\documentclass[aps,prb,twocolumn,showpacs,amsmath,amssymb,superscriptaddress,floatfix]{revtex4-1}
\usepackage{graphicx}
\usepackage{dcolumn}
\usepackage{bm}
\usepackage{amsfonts}
\usepackage{amsmath}
\usepackage{ulem}
\usepackage{color}
\usepackage{hyperref}

\begin{document}

\title{Noise and current correlations in tunnel junctions of Quantum Spin Hall edge states}

\author{Fabrizio Dolcini}
\email{fabrizio.dolcini@polito.it}

\affiliation{Dipartimento di Scienza Applicata e Tecnologia del Politecnico di Torino, I-10129 Torino, Italy}

\affiliation{CNR-SPIN, Monte S.Angelo - via Cinthia, I-80126 Napoli, Italy}

\begin{abstract}
The edge channels of two-dimensional topological systems  are protected from elastic reflection and are   noiseless at low temperature. Yet, noise and cross-correlations can be induced when electron waves   partly transmit  to the opposite  edge via  tunneling through a constriction. 
In particular, in a quantum spin Hall (QSH) system tunnelling occurs via both  spin-preserving ($p$) and  spin-flipping~($f$) processes, each fulfilling time-reversal symmetry.
We investigate the current  correlations of a four-terminal QSH  setup in the presence of a tunneling region, both at equilibrium and out-of-equilibrium. We find that, although  $p$ and $f$ processes do not commute and the generic current correlation depends on both, under appropriate conditions a direct  detection of two types of partition noise is possible. In particular, while the spin-preserving partitioning can   be probed  for any arbitrary tunnel junction with a specific configuration of terminal biases, the spin-flipping partitioning can be directly detected  only under suitably designed setups and conditions. We describe two setups where these conditions can be fulfilled, and both types of partitioning can be detected and controlled.  
\end{abstract}


\maketitle

\section{Introduction}
The spectrum of the current-current correlations, also called noise, is known to provide extremely useful physical insights about electronic transport. At equilibrium, the Johnson-Nyquist noise originates from thermal fluctuations and  is closely related to the conductance. However, when the system is driven out of equilibrium by an applied bias $V$, an additional noise source arises that survives even at low temperatures:  the shot noise. For each conducting channel characterized by reflection and transmission coefficients $R$ and $T=1-R$ , the shot noise exhibits the well known partitioning expression proportional to $R \,T$, which is the hallmark of the discreteness of electronic charge and is smaller than  Poissonian noise.~\cite{lesovik_1989,buttiker_prl_1990} 

In quantum cavities or in diffusive quantum wires many channels are present and shot noise originates from backscattering off impurities or barriers that partly reflect and partly transmit electron waves. In contrast,  two-dimensional (2D) topological systems are  characterized by an insulating bulk and  a limited number of conducting channels  flowing at the edges of a quantum well. Because these edge states are topologically protected from impurity scattering, they behave as perfectly conducting noiseless channels at low temperatures. However, if electron tunneling between two opposite edges is induced, e.g. by realizing a constriction in the quantum well, the topological protection   is locally lost, and electrons can be partly ``reflected'' from one channel to the opposite one. Electron  shot noise and cross correlations thus  arise in the multi-terminal setup, providing useful physical information. In Quantum Hall (QH) systems, for instance, shot noise  has been analyzed since long, leading to observe sub-Poissonian correlations~\cite{washburn_1991},   fermion antibunching~\cite{schonenberger_1999} and even the fractional charge underlying  the fractional QH  regime.~\cite{depicciotto_1997}

In contrast, in the recently discovered Quantum Spin Hall (QSH) systems,~\cite{kane-mele_2005,bernevig_science_2006,bernevig_prl_2006,qi-zhang_2011} noise measurements are lacking,  so far. These 2D topological systems, realized in HgTe/CdTe~\cite{konig_2006,molenkamp-zhang_jpsj,roth_2009,brune_2012}
or in InAs/GaSb
 quantum wells~\cite{liu-zhang_2008,knez_2007,knez_2014,spanton_2014}, exhibit  {\it helical} edge states, where the group velocity is locked to the spin orientation, so that the two counter-propagating modes along a given edge exhibit opposite spin orientations.   
As far as the theoretical analysis of noise in QSH systems is concerned,   most studies have focussed on the effect of electron interaction on the current-current correlations, within the helical Luttinger liquid model in the presence of a point-like constriction  in the QSH bar.~\cite{schmidt_2011,souquet_2012,lee_2012,posske_2014} However,  at the moment the experimental evidence of helical Luttinger regime is quite limited.~\cite{du_2015}  

In order to boost experimental research on noise in QSH systems, further motivations that do not specifically rely on the role of interaction are thus needed. 
So far, proposals are limited to the case where  a magnetic field is applied to the QSH bar, pointing out the appearance of a noise peak.~\cite{buttiker_2013} 
However,  the very existence of topological helical states 
boils down to spin-orbit coupling,~\cite{kane-mele_2005,bernevig_science_2006,bernevig_prl_2006,qi-zhang_2011} which should reveal its signatures on electron noise even when time-reversal symmetry is not broken. This is known to be the case already for topologically trivial conductors, where both Rashba and Dresselhaus spin-orbit couplings have been proven to strongly affect the noise, with interesting implications for spintronics.~\cite{loss_2002,beenakker_2006,nikolic_2007,zu_2007}  For the helical states, a constriction in the QSH bar causes a local wavefunction overlap~\cite{zhou,richter} that is known to yield two types of tunneling processes.~\cite{zhang-PRL,teo,richter,trauz-recher,dolcini2011,citro-romeo,citro-sassetti,sassetti-cavaliere-2013,moskalets_2013,bercioux_2013,sternativo1}
The first type is the customary spin-preserving ($p$) process, where an electron tunnels across the junction maintaining its spin orientation, and thereby reversing its group velocity. The second type is less conventional and is a spin-flipping ($f$) process, where a tunneling electron reverses its spin orientation   maintaining its group velocity: it mainly originates from the interplay between bulk-inversion asymmetry and wavefunction overlap,  even in absence of magnetic coupling.  
In fact   these  two types of tunneling fulfill time-reversal symmetry, and identify two different reflection coefficients $R_{p}$ and $R_{f}$, whose control and analysis is a crucial issue for helical state based spintronics.~\cite{moore_2010,qi-zhang_2010,murakami_2014}

As far as average electron currents are concerned, these two processes affect differently the redistribution of the injected electrons onto the various terminals.~\cite{sternativo1,sternativo2} Natural questions now arise about the current fluctuations. What is their behavior, and how do they depend on the configurations of the voltage biases applied to the various terminals~? And, notably, is there an operative way to observe two types of shot noise, namely a spin-preserving partitioning ($\propto R_{p} T_{p}$) and a spin-flipping partitioning ($\propto R_{f} T_{f}$),   at least under some conditions~? This article addresses  these questions.

The paper is organized as follows: in Sec.~\ref{sec-II}, we describe the model   for a four-terminal setup with a tunnel junction between helical edge states. In Sec.~\ref{sec-III} we derive the expression for the current-current correlation matrix, first for the equilibrium case and then focussing on the shot noise regime, where various voltage bias configurations are analyzed. The general condition for detecting the two types of shot noise are determined. In Sec.~\ref{sec-IV} we present some explicit results about noise for two specific setups that fulfill the found conditions. Finally, in Sec.~\ref{sec-V} we summarize our results and draw some conclusions.

 
\section{Model}
\label{sec-II}
We consider a four-terminal QSH bar where, for definiteness, we assume that along the top edge right movers are characterized by spin-$\uparrow$  and left  movers by spin-$\downarrow$, while the opposite spin orientations   characterize the bottom edge. 
A  constriction is realized over the central region of the QSH bar, allowing transversal electron tunneling between the four edge states. 
The injection of helical states into the tunneling region is controlled by four metallic terminals, which we shall label  in clockwise order,  from the bottom left corner (terminal~1) to the bottom right corner (terminal~4) of the setup. 
Two gate electrodes, applied at the  sides of the constriction, enable one to shift the chemical potential of the edge states. We adopt the model discussed in detail in  Refs.[\onlinecite{sternativo1}] and [\onlinecite{sternativo2}]. Here below we shall briefly summarize the main ingredients that are needed for the present analysis of the   current-current correlations.
\\

The electron field operators  
$\Psi_{R\uparrow}(x) ,
\Psi_{L\downarrow}(x)$ are utilized to account for the helical  states of  the   top edge, and $\Psi_{R\downarrow}(x) ,
\Psi_{L\uparrow}(x)
$ for   the ones in the bottom edge, with  $x$ denoting  the longitudinal QSH bar direction. The Dirac Hamiltonian describing the uncoupled edges reads~\cite{kane-mele_2005,bernevig_science_2006,bernevig_prl_2006}
\begin{equation}
\hat{\mathcal{H}}_0 = -i    \hbar v_{F} \hspace{-0.35cm} \sum_{\alpha= R/L=\pm} \hspace{-0.2cm}\alpha \sum_{\sigma=\uparrow, \downarrow}\int   \! dx     :   \Psi^{\dagger}_{\alpha \sigma}(x)\, \partial_x \Psi^{}_{\alpha \sigma}(x)    :                 
\label{H0}
\end{equation}
where $v_F$ is the Fermi velocity, $\alpha=R/L=\pm$ denotes the group velocity for right- and left- movers, respectively, $\sigma=\uparrow,\downarrow$  the spin component, and $: \, \, \, :$ the normal ordering with respect to the Dirac point. In the central region of the setup, the two edge states are coupled by electron tunneling. The  tunneling Hamiltonian that preserves time-reversal symmetry consists of two terms~\cite{zhang-PRL,teo,richter,trauz-recher,dolcini2011,citro-romeo,citro-sassetti}, $\hat{\mathcal{H}}_{\rm tun}=\hat{\mathcal{H}}_{\rm tun}^p+\hat{\mathcal{H}}_{\rm tun}^f$, where
\begin{eqnarray}
\hat{\mathcal{H}}_{\rm tun}^p  &=& \! \!  \displaystyle \sum_{\sigma=\uparrow, \downarrow}\! \int   \! dx      \left( \Gamma_{p}^{}(x) \,  \Psi^{\dagger}_{L \sigma}(x)\, \Psi^{}_{R \sigma}(x)    +   \mbox{H.c.} \right) \, \;  
\label{Htunp} \\
\hat{\mathcal{H}}_{\rm tun}^f &=&    \hspace{-0.4cm}\sum_{\alpha=R/L=\pm} \hspace{-0.3cm} \alpha \int   \! dx    \,  \left(   \Gamma_{f}^{}(x) \, \,  \Psi^{\dagger}_{\alpha \downarrow}(x)\, \Psi^{}_{\alpha \uparrow}(x)   \, +     \mbox{H.c.} \right)   ,  \, \label{Htunf}  
\end{eqnarray}
describe spin-preserving tunneling (where the group velocity is reversed), and spin-flipping tunneling (where the group velocity is preserved), respectively. 
Here $\Gamma_p(x)$ and $\Gamma_f(x)$ are complex tunneling amplitudes, whose magnitude is determined by the local transversal width   of the junction along the longitudinal direction $x$. We retain  only contributions from transversal tunneling, for they are proven to be largely dominant.~\cite{dolcetto-sassetti}
In addition, we  consider  external potentials $V_{T}$ and $V_{B}$ that couple to the two edges 
\begin{eqnarray}
\hat{\mathcal{U}}  &=&  \displaystyle  \! \int    \! dx   \left[  \, eV_{T}(x) \left(   \hat{\rho}_{R \uparrow}(x)\,     \,+  \hat{\rho}_{L \downarrow}(x)\,  \right) \right. \label{U}  \\
  & &  \left. \displaystyle  \hspace{0.5cm} +eV_{B}(x) \left(   \hat{\rho}_{R \downarrow}(x)    \,+ \,  \hat{\rho}_{L \uparrow}(x)   \right) \right] \quad,
\nonumber 
\end{eqnarray}
where $
\hat{\rho}_{\alpha \sigma}(x)=\, :\Psi^\dagger_{\alpha \sigma}(x) \Psi^{}_{\alpha \sigma}(x):
$
is the electron chiral density. Notice that the term (\ref{U}), alone, would simply shift the Dirac cone given by Eq.(\ref{H0}), with no effect on transport. However, because of the presence of the tunneling terms (\ref{Htunp}) and (\ref{Htunf}), $\hat{\mathcal{U}}$ does affect the current redistribution in the setup. 
Indeed, as we shall see in the following, the two types of partitioning can be controlled by the two combinations $V_{p/f}=(V_{T}\pm V_{B})/2$, which we shall term the charge and spin gate, respectively. 
The full Hamiltonian  thus reads as
$
\hat{\mathcal{H}} =\hat{\mathcal{H}}_{0}   \, + \,   
\hat{\mathcal{H}}_{\rm tun}   \, + \,
\hat{\mathcal{U}}       
$.\\

The  longitudinal profile of the tunneling amplitudes $\Gamma_{p}(x) \, , \Gamma_{f}(x)$  and of the side gate potentials $V_{T}(x),V_{B}(x)$ -- or  equivalently the charge and spin gates $V_{p}(x),V_{f}(x)$-- characterize the  tunnel setup. Specific cases  will be discussed Sec.\ref{sec-IV}. Here, without loss of generality  we only assume that, sufficiently  far away from the central region, tunneling and potential terms vanish. Explicitly, denoting by $x_0$ and $x_f$   the two extremal coordinates of the central region, one has $\Gamma_{\nu}(x), V_{\nu}(x) \neq 0$ ($\nu=p,f$) only for  $x_0 \le x  \le x_f$. Then, the stationary solutions $\Psi(x,t)=\int dE e^{-i E t/\hbar} \Psi_E(x)$ of the equation $i\hbar \, \partial_t  \Psi =[ \Psi  \,, \hat{\mathcal{H}}]$ for the four-component field operator $
\Psi(x) =\left( 
\Psi_{R\uparrow}(x) ,
\Psi_{L\uparrow}(x) ,
\Psi_{R\downarrow}(x) ,
\Psi_{L\downarrow}(x) 
\right)^T
$ acquire the asymptotic form $\Psi_E(x \le x_0)=  (\sigma_0 \otimes e^{i \tau_z k_Ex}  ) \,\left(   {a}_{R\uparrow}, {b}_{L\uparrow} , {a}_{R\downarrow},  {b}_{L\downarrow}  \right)^T  /\sqrt{2\pi \hbar v_F}$
and $\Psi_E(x \ge x_f)=   (\sigma_0 \otimes e^{i \tau_z k_Ex})   \,\left(   {b}_{R\uparrow}, {a}_{L\uparrow} , {b}_{R\downarrow},  {a}_{L\downarrow} \right)^T / \sqrt{2\pi \hbar v_F}$,
where $a_{\alpha \sigma}$ and $b_{\alpha \sigma}$ denote operators for incoming and outgoing states, respectively, and $k_E=E/\hbar v_F$.

The four-terminal scattering matrix~$\mathsf{S}$ relating outgoing  to incoming operators,~\cite{buttiker-review}
\begin{equation}
\left( \begin{array}{c}  {b}_{L\uparrow}\\  {b}_{L\downarrow} \\  {b}_{R\uparrow}\\  {b}_{R\downarrow} \end{array} \right) \, = \, {\mathsf{S}} \, \left( \begin{array}{c}  {a}_{R\downarrow}\\  {a}_{R\uparrow} \\  {a}_{L\downarrow}\\  {a}_{L\uparrow} \end{array} \right) \quad,
\end{equation}
can be proven  to acquire  the expression~\cite{sternativo1}
\begin{equation}
{\mathsf{S}}= \left( \begin{array}{cccc}
0 & r_{p}  &  t_{p} \, r^*_{f}  & t_{p} \, t^*_{f}  \\ & & & \\
 r_{p} & 0 & t_{p} \,t_{f} & -t_{p} \, r_{f} \\ & & & \\
- t_{p} \, r^*_{f} & t_{p} \, t_{f} & 0 & r^\prime_{p}  \\ & & & \\
t_{p} \, t^*_{f} & t_{p} \, r_{f} & r^\prime_{p} & 0
\end{array}\right) \quad,\label{S-res}
\end{equation} 
where $r_p$ and $t_p$ (with $r^\prime_{p}=-r^*_{p} \,  t_{p}/t^*_{p}$) are spin-preserving reflection and transmission amplitudes, as they depend only on the tunneling amplitude $\Gamma_{p}$ and on the charge  voltage $V_{p}$. Similarly, $r_{f}$ and $t_{f}$ are spin-flipping  reflection and transmission amplitudes, for they depend  on the spin-flipping tunneling amplitude $\Gamma_{f}$ and on the spin gate voltage $V_{f}$ only.

Notably, in Eq.(\ref{S-res})  each entry is {\it factorized} into the product of ${p}$- and ${f}$-terms. As shown in Ref.[\onlinecite{sternativo1}], such simple factorized form, which holds despite the two tunneling terms~(\ref{Htunp}) and (\ref{Htunf}) do not  commute,  enables one to operatively determine the reflection coefficients $R_p=|r_p|^2$ and $R_f=|r_f|^2$ related to the two types of tunnel processes, through transconductance measurements. If, for instance, terminal 2 is biased and currents are measured in the other terminals, $R_{p}$ is extracted from the conductance $|{\rm G}_{12}|= (e^2/h) R_p$ in terminal 1. Then, $R_{f}$ is determined from the conductance $|{\rm G}_{42}|= (e^2/h) R_f T_p$ in terminal  4, where  $T_{p}=1-R_{p}$. Equivalently $T_{f}=1-R_{f}$ is  determined from $|{\rm G}_{32}|= (e^2/h) T_f T_p$ in terminal 3.  


\section{Current-current correlations}
\label{sec-III}
The frequency spectrum of the  current-current correlations between any two terminals $i,j=1,\ldots 4$ is defined as
\begin{equation}
{\rm P}_{ij}(\omega) \doteq \int_{-\infty}^{+\infty} e^{i \omega t}\left( \langle \hat{I}_i(t)\, \hat{I}_j(0) \rangle - \langle \hat{I}_i(t)\rangle\, \langle \hat{I}_j(0) \rangle\right)\, dt \quad,\label{Pij-def}
\end{equation}
where the current  operator  $\hat{I}_i$ in each terminal is defined as positive when it is incoming from that terminal to the scattering region, according to the customary convention for multi-terminal setups.~\cite{buttiker_prb_1992,buttiker_njp_2007}  Following the clockwise order of the terminals from the bottom left corner of the setup, one has 
$\hat{I}_1 \doteq   e v_F  ( \hat{\rho}_{R\downarrow}  -\hat{\rho}_{L\uparrow})$, 
$\hat{I}_2 \doteq   e v_F  ( \hat{\rho}_{R\uparrow} -\hat{\rho}_{L\downarrow} )$,
$\hat{I}_3 \doteq   e v_F  ( \hat{\rho}_{L\downarrow}  -\hat{\rho}_{R\uparrow} )$, and 
$\hat{I}_4 \doteq   e v_F ( \hat{\rho}_{L\uparrow}  -\hat{\rho}_{R\downarrow} )$. 
By applying  standard methods of Scattering Matrix formalism,~\cite{buttiker-review}  one can compute the current correlations (\ref{Pij-def}) for the four-terminal setup, in terms of the Fermi distributions $f_i(E)=\{1+\exp[(E-\mu_i)/k_B T]\}^{-1}$  ($i=1,\ldots 4$) of the  reservoirs, characterized by a temperature $T$ and a chemical potential~$\mu_i$. The obtained 16 combinations of current correlations ${\rm P}_{ij}$ in Eq.(\ref{Pij-def}) are thus  cast into a $4 \times 4$ current correlation matrix ${\rm P}$. We shall henceforth focus on the customary low-frequency case, and set $\omega=0$. \\

\subsection{Johnson-Nyquist noise}  We  start by analyzing the Johnson-Nyquist noise, i.e. the noise arising when the setup is at thermal equilibrium at a temperature $T$, and all terminals are characterized by the same distribution $f_0$ with the same chemical potential $\mu_i \equiv E_F$, \, $\forall i=1,\ldots 4$. Then, a straightforward calculation enables one to find
\begin{equation}
{\rm P}_{ij}^{\rm JN}= 2 \int {\rm G}_{ij}(E) \, f_0(E)(1-f_0(E)) \,dE \quad,\label{JN}
\end{equation}
where  ${\rm G}_{ij}=(e^2/h)(\delta_{ij}-|\mathsf{S}_{ij}|^2)$ denote  the entries of the conductance matrix~${\rm G}$, describing the current flowing through  terminal $i$ when a small voltage bias is applied to terminal $j$,  obtained from the Scattering matrix~(\ref{S-res}).  In particular, when the transmission coefficients $T_p$ and $T_f$ vary over energy ranges bigger than thermal energy $k_B T$, the Johnson-Nyquist current correlations (\ref{JN}) acquires the simple form ${\rm P}_{ij}^{\rm JN}= 2 k_B T  {\rm G}_{ij}(E_F)$, which shows that at thermal equilibrium the current correlation matrix ${\rm P}$ is essentially given by the transconductance matrix ${\rm G}$, in accordance with the fluctuation-dissipation theorem. Explicitly, the Johnson-Nyquist   current-current correlation matrix reads as
{\small \begin{equation}
{\rm P}^{\rm JN}=\frac{2e^2}{h} k_B T \left( \begin{array}{cccc}
1 & -R_{p}  &  -T_{p} R_{f}  & -T_{p} \, T_{f}  \\ & & & \\
-R_{p} & 1 & -T_{p} T_{f} & -T_{p} \, R_{f} \\ & & & \\
-T_{p} R_{f} & -T_{p} T_{f} & 1 & -R_{p}  \\ & & & \\
-T_{p} \, T_{f} & -T_{p} \,  R_{f} & -R_{p} & 1
\end{array}\right)
\label{P-JN} .
\end{equation}
}  
Here the diagonal terms represent the local noise, i.e. the current fluctuations within the same terminal, whereas the off-diagonal terms  ${\rm P}_{i \neq j}$ describe the cross correlations of currents at different terminals, and exhibit the customary negative sign due to the fermion antibunching.~\cite{buttiker_prb_1992,buttiker_njp_2007}  Notice that the sum of all contributions along each row or column vanishes, due to the conservation of the total current.
  
The first consequence of Eq.(\ref{P-JN}) is that, in a QSH setup, the Johnson-Nyquist local noise  in each terminal is   independent of the properties of tunnel region, and is simply given by the  quantum of conductance times twice the thermal energy, ${\rm P}_{ii}^{\rm JN}=2k_B T e^2/h$. This universality of the local equilibrium noise boils down to the topological protection of helical edge states, and does not occur in ordinary time-reversal invariant systems, where backscattering onto the injection terminal can occur.~\cite{buttiker_prb_1992} Secondly,   Eq.(\ref{P-JN}) also shows that  the Johnson-Nyquist cross correlations ${\rm P}_{i\neq j}^{\rm JN}$ provide  a straightforward way to operatively extract the reflection coefficients $R_p$ and $R_f$ ascribed to the two types of tunneling processes. Explicitly, one can first gain  $R_p$ from the cross correlations ${\rm P}_{12}^{\rm JN}$, and thereby extract $R_f$ from the cross correlations  ${\rm P}_{42}^{\rm JN}$,   or equivalently $T_f=1-R_{f}$ from ${\rm P}_{32}^{\rm JN}$.
This prescription is alternative and equivalent to transconductance detection method described in Ref.[\onlinecite{sternativo1}]. Notice, in passing, that the prefactor $2 k_B T$ relating Johnson-Nyquist noise and transconductance in the  relation ${\rm P}_{ij}^{\rm JN}= 2 k_B T  {\rm G}_{ij}(E_F)$ is one half of the customary prefactor $4 k_B T$ found in  multi-terminal transport of conventional materials,~\cite{buttiker-review,buttiker_prb_1992,buttiker_njp_2007}  again due to the   helical nature of the edge states, where only  half of the channels are involved. \\

\subsection{Shot noise and voltage bias configurations}  
\label{shot}
Let us now consider the out-of-equilibrium noise by letting the chemical potential $\mu_i=E_F+eV_i$ of each lead $i=1\ldots 4$ deviate  from the equilibrium level $E_F$ by a bias voltage $V_i$. Because of the presence of four terminals, various   configurations of applied biases are  possible and the out-of-equilibrium current correlations do depend on them. In view of identifying  different types of noise terms, we shall focus on two inequivalent configurations, in which two of the four terminals are biased to $+V/2$ and other two terminals are biased to  $-V/2$. For each configuration, we shall focus on the shot noise regime, $eV \gg k_B T$, where thermal contributions to noise are negligible.   

\subsubsection{charge-bias configuration} In this configuration the terminals are biased as  $V_1=V_2=V/2$ and $V_3=V_4=-V/2$, and     the noise matrix exhibits a fully factorized form, 
\begin{equation}
{\rm P}^{\rm CB} = \frac{e^2}{h}  {\small \left( \begin{array}{cccc}
1 &   0  &    -  T_{f}  &    -  R_{f}  \\   & & & \\
0 &   1 &   -  R_{f} &   -  T_{f}  \\  & & &\\
 -  T_{f}   &  -  R_{f}   & 1   & 0 \\ & & & \\
- R_{f} &    -  T_{f} &   0   & 1
\end{array}\right)}\! \int_{-\frac{eV}{2}}^{\frac{eV}{2}} \! \! \! dE \, R_{p}  T_{p}   \,,
\label{shot-CB}\vspace*{1cm}
\end{equation}
where a  partition noise expression  $R_{p} T_{p}$  ascribed to purely spin-preserving tunneling processes singles out and multiplies an energy-independent matrix that depends on the spin-flipping processes only. In particular, since the diagonal  local noise entries are insensitive to spin-flipping processes occurring in the junction, the spin-preserving shot noise ${\rm P}_{p}$    can be operatively  identified as the local noise in the charge-bias configuration,
\begin{eqnarray}
{\rm P}_{p} &\doteq &{\rm P}^{\rm CB}_{ii} =\frac{e^2}{h}  \int_{-\frac{eV}{2}}^{\frac{eV}{2}} \! \! \! dE \, R_{p}  T_{p} \hspace{1cm} \forall i=1,\ldots 4 \, \, \, \label{P-p-def} \\
& \simeq & \frac{e^3}{h}  V  \, R_{p}  T_{p} \quad,\label{P-p-def-2}
\end{eqnarray}
where the second line holds when the applied bias $V$ is small compared to the typical range of variation of the reflection coefficient $R_{p}$.

As far as the off-diagonal cross correlations in Eq.(\ref{shot-CB}) are concerned, they vanish between   terminals characterised by the same voltage,  ${\rm P}_{12}={\rm P}_{34}=0$, whereas the  correlations ${\rm P}_{13}$ and ${\rm P}_{14}$   describe a splitting of the spin-preserving partition noise $R_{p} T_{p}$ into two components, proportional to the spin-flipping transmission and reflection $T_f$ and $R_f$, respectively. In the charge-bias configuration, spin-flipping tunneling processes act  as a `splitter' for the spin-preserving partition noise $R_{p} T_{p}$. Notice that   the matrix (\ref{shot-CB}) can be regarded to as consisting of four $2 \times 2$ blocks. Formally, summing up the entries of each   block corresponds to  merging   the signals of the terminals characterised by the same bias --(1,2) and (3,4)--, and to   recover an  effective two-terminal setup of a spinful quantum wire, where the tunnel region plays the role of a barrier.  In doing that, however, any trace of spin-flipping processes would be completely lost, since $R_{f}+T_{f}=1$. In that respect, the four-terminal QSH setup corresponds to a spin-resolved current-current correlation measurement, where spin-flipping processes do matter.

\subsubsection{spin-bias configuration} 
This configuration is defined by the following  biases  $V_2=V_4=V/2$ and $V_1=V_3=-V/2$. We find for the  noise matrix  
\begin{widetext}
\begin{equation}
{\rm P}^{\rm SB} = \frac{e^2}{h}  \int_{-\frac{eV}{2}}^{\frac{eV}{2}} \! \!  \, {\small \left( \begin{array}{ccccccc}
(1-T_{p} R_{f})T_{p} R_{f}  & & - T_{p}^2 R_{f} T_{f}   & & 0    & &  -  R_{p} T_{p} R_{f}  \\ & & & \\
- T_{p}^2 R_{f} T_{f} & &(1-T_{p} R_{f})T_{p} R_{f} & & -  R_{p} T_{p} R_{f} & & 0 \\ & & & & &  &  \\
0  & &  - R_{p} T_{p}  R_{f}  & &  (1-T_{p} R_{f})T_{p} R_{f}  & & -T_{p}^2 R_{f} T_{f} \\ & &  & &  & & \\
-   R_{p} T_{p} R_{f}   & & 0 & & -T_{p}^2 R_{f} T_{f}  & & (1-T_{p} R_{f})T_{p} R_{f}
\end{array}\right)}\,   dE
\label{shot-SB} \vspace*{1cm}
\end{equation}
 
\end{widetext}  
and no overall factorisation between spin-preserving and spin-flipping terms occurs. The diagonal local noise entries, for instance,  exhibit a partition form $T_{eff}(1-T_{eff})$ where the effective transmission coefficient $T_{eff}=T_{p} R_{f}$  mixes both sectors. The cross correlations are characterized by other mixed combinations of ${p}$- and ${f}$-processes.  \\\\

The inspection of Eqs.(\ref{shot-CB})-(\ref{shot-SB}) enables us to make the following remarks about out-of-equilibrium current-current correlations. In the first instance, when no tunneling occurs ($R_{p}=R_{f}=0$), all entries of Eqs.(\ref{shot-CB})-(\ref{shot-SB}) vanish, showing that uncoupled helical edges are noiseless at low temperatures, as a result of topological protection from backscattering. Secondly, when tunneling is present, a difference can be noticed between spin-preserving and spin-flipping noise.
Notably, while the partition noise $T_{p} R_{p}$ for spin-preserving tunneling directly appears as the local noise in the charge-bias configuration (\ref{shot-CB}), a spin-flipping partition noise $T_{f} R_{f}$  is absent in both configurations Eqs.(\ref{shot-CB}) and (\ref{shot-SB}). It can be easily proven that this lack also occurs in any other bias configuration, as a direct consequence of the Scattering matrix (\ref{S-res}): while $\mathsf{S}$  does exhibit  entries $\mathsf{S}_{12}$ and $\mathsf{S}_{34}$ that depend on ${p}$-terms only (i.e. the entries related to terminals located on the same side of the tunnel region), no entry that depends  on {\it purely} ${f}$-terms appears. Because the shot noise terms are essentially given by products of  scattering matrix entries, a partition noise $T_{f} R_{f}$ cannot occur, and its straightforward observability seems jeopardised.

However, the following remark can be made: if one could ideally ``switch-off'' the spin-preserving processes and set $R_{p}=0$, the shot noise would only originate from spin-flipping processes. Then,  the current-current correlation matrix in the spin-bias configuration (\ref{shot-SB}) would exhibit a simple block-diagonal form, 
\begin{equation}
{\rm P}^{\rm SB} \rightarrow \frac{e^3}{h}  R_{f} T_{f} \, V  \! \!  \, {\small \left( \begin{array}{ccccccc}
1  & & - 1   & & 0    & &  0 \\ & & & \\
- 1 & & 1 & & 0 & & 0 \\ & & & & &  &  \\
0  & &    0  & &  1  & & -1\\ & &  & &  & & \\
 0 & & 0 & & -1 & & 1
\end{array}\right)} \quad,    
\label{shot-SB-bis} \vspace*{1cm}
\end{equation}
where both the local noise and the cross correlations between the terminals (1,2) and (3,4) are characterized by 
the partitioning $R_{f} T_{f}$. The  spin-flipping shot noise can thus be operatively defined as the current correlation in the spin-bias configuration under the constraint that $R_p =0$, 
\begin{equation}
{\rm P}_{f}=  {\rm P}^{\rm SB}_{ii}=-{\rm P}^{\rm SB}_{12}=-{\rm P}^{\rm SB} _{34}= \frac{e^3}{h}  V \,  R_{f} T_{f} \hspace{0.5cm} \mbox{$(R_p =0)$.} \label{P-f-def}
\end{equation}

The physical meaning of the condition $R_p =0$ can be easily understood by recalling the close relation between the spin-preserving reflection $R_{p}$ and the transconductance $|{\rm G}_{12}|=(e^2/h) R_{p}$: when the terminal 2 is biased, $R_{p}=0$ implies that no current is detected in terminal 1, i.e. the spin-preserving tunneling,  acting as a back-scattering into the opposite edge, vanishes. Since for weak tunneling $R_{p} = \mathcal{O}(|\Gamma_{p}|^2)$, one could naively think of fulfilling the above condition simply by completely removing the constriction, or by making it extremely weak. However, since  one typically has $|\Gamma_f|< |\Gamma_p|$, that would also imply a vanishing of the spin-flippping tunneling amplitude, making also $R_{f} \rightarrow 0$, and thereby hindering the possibility to observe the spin-flipping partitioning $T_{f} R_{f}$. It thus seems that the condition $R_{p}=0$ is too restrictive and   a direct detection of spin-flipping shot noise is impossible. 
In the next section, we shall show that this is not necessarily the case.

\section{Spin-preserving and spin-flipping partitioning}
\label{sec-IV}
In the previous section we have outlined the general properties of the current-current correlations of any four-terminal QSH setup where tunneling region is present. In particular, in Sec.\ref{shot} we have focussed on out-of-equilibrium conditions, and on the possibility to probe spin-preserving  partitioning $R_{p} T_{p}$ and  spin-flipping partitioning $R_{f} T_{f}$: While the former can  be extracted directly in any type of tunnel setup by performing a local noise measurement in the charge-bias configuration, the latter can be directly probed in the spin-bias configuration only provided the condition $R_{p}=0$ is satisfied. As observed above, such condition is not fulfilled in a generic setup, since the ${p}$-tunnel processes cannot be removed while keeping $f$-tunneling processes.

In this section we describe two specific proposals for setups where such drawback can be overcome and both spin-preserving and spin-flipping partitioning can be directly probed. The underlying idea is that, although ${p}$-tunnel processes cannot be removed, they can be made inactive in setups where  the reflection coefficient $R_{p}$ is energy-dependent and it can be   tuned to extremely small values. The two setups that we shall consider are an extended tunnel junction (ETJ)  and a double quantum point contact (DQPC),   which act as an energy-filter and an electron wave  interferometer, respectively. In both cases, two side  gate voltages $V_{T}$ and $V_{B}$ -- located along the junction length in the ETJ setup and between the two QPCs in the DQPC setup-- can be suitably applied  to realise the condition $R_{p} \simeq 0$, necessary to observe the spin-flipping partition noise. To this purpose, we shall exploit the property mentioned in Sec.~\ref{sec-II} that   $R_{p}$ depends only on the combination $V_{p}=(V_{T}+V_{B})/2$, while $R_{f}$ depends only on the difference $V_{f}=(V_{T}-V_{B})/2$. 
Without loss of generality, we shall assume in the following that the equilibrium Fermi level, when no gate and bias is applied, is located at the Dirac point, $E_F=0$.

\subsection{The Extended tunnel Junction (ETJ)}  
\begin{figure} 
\centering
\includegraphics[width=\columnwidth,clip]{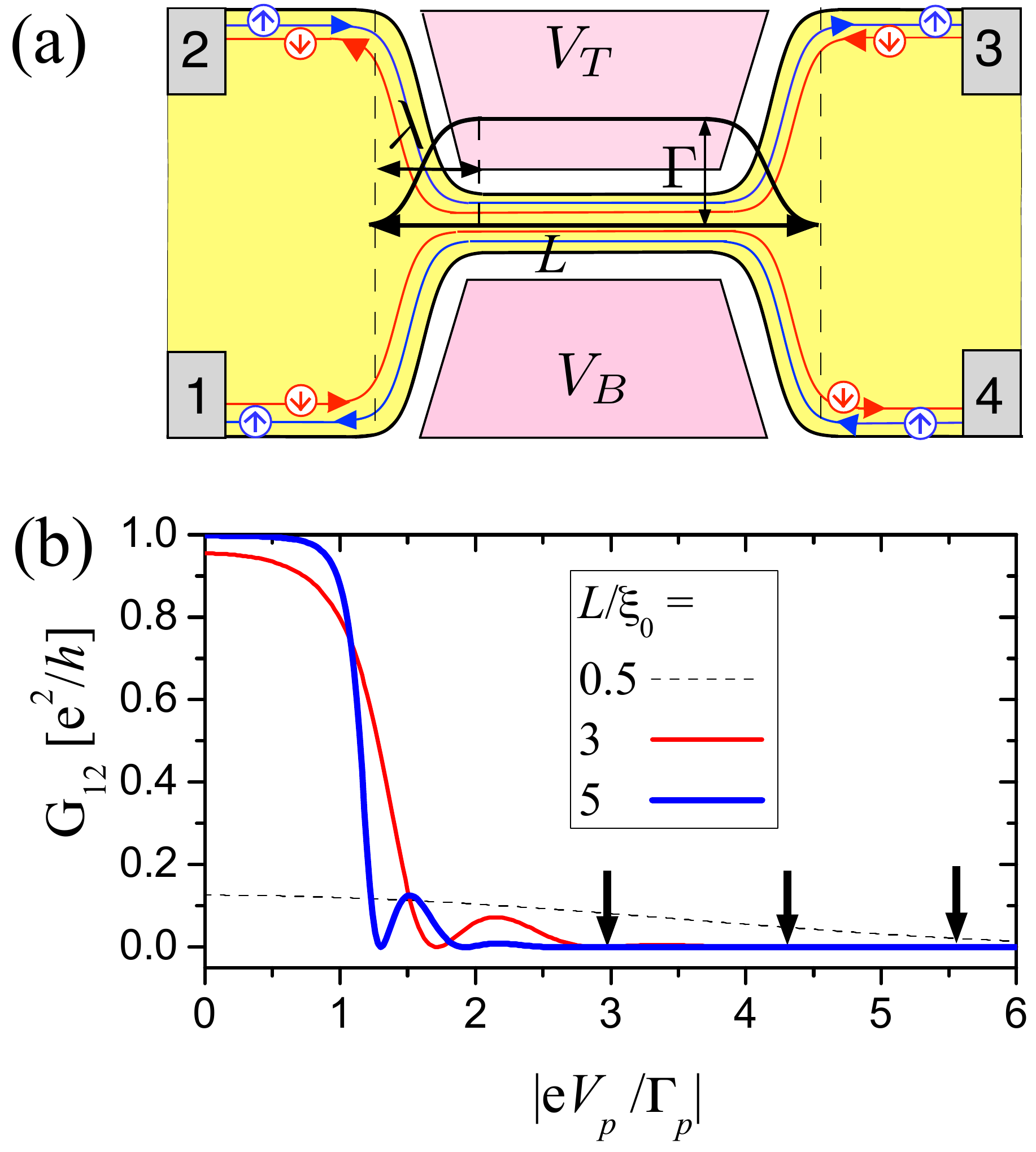}
\caption{(Color online) (a) Sketch of the  extended tunnel junction setup, characterized by a total length $L$, where the   amplitudes for spin-preserving and spin-flipping tunnel processes increase from 0 to the bulk values $|\Gamma_{p}|$ and $|\Gamma_{f}|$,  over a lengthscale $\lambda$. Two side gate voltages $V_{T}$ and $V_{B}$ are applied across the tunnel junction.
(b) The transconductance $|{\rm G}_{12}|=(e^2/h) R_p$ is plotted  as a function of the charge gate voltage $V_{p}=(V_{T}+V_{B})/2$, for three different ratios of the junction length $L$ to the lengthscale $\xi_0=\hbar v_F/|\Gamma_p|$, and for $\lambda=0.25 L$. While for short junctions the reflection is always small and weakly dependent on $V_{p}$, for extended junctions $|{\rm G}_{12}|$ drops from the quantum of resistance to 0, within an energy window around the bulk spin-preserving tunneling amplitude $|\Gamma_p|$. The vertical arrows indicate the regime where the condition $R_{p} \simeq 0$ for revealing the spin-flipping partition noise is fulfilled (see Fig.\ref{Fig-ETJ-noise}(b)).}
\label{Fig-ETJ-setup}
\end{figure}

The first setup, depicted in   Fig.\ref{Fig-ETJ-setup}(a), consists of a long constriction in the QSHE quantum well, leading to  tunneling amplitudes $|\Gamma_{p,f}(x)|$ that increase from 0  away from the tunnel region to a  value $|\Gamma_{p,f}|$ in the `bulk'  of the junction, over a length $\lambda$. Because  spin-preserving tunneling induces backscattering across the other edge, the tunnel junction effectively acts as a barrier. For a short junction, $L < \xi_0$ with $\xi_0=\hbar v_F/|\Gamma_p|$, the `barrier' weakly reflects electrons from terminal 2 into terminal 1.
As a consequence, the transconductance $|{\rm G}_{12}|=(e^2/h) R_p$ is small for $E < |\Gamma_p|$ and tends to 0 for $E \gg |\Gamma_p|$ with a very smooth crossover (thin dashed curve in Fig.\ref{Fig-ETJ-setup}(b)), which makes it more difficult to identify. 
In contrast, in an extended tunnel junction, $L > \xi_0$,  the electronic waves  decay along the junction length $L$ over a localisation length $\xi_E=\xi_0 \sqrt{1-( E/|\Gamma_p|)^2}$ for $E<|\Gamma_p|$, while  electrons behave as propagating  waves with   transmission   close to 1 for $E>|\Gamma_p|$. An ETJ thus behaves as a high-pass energy filter, where the switching between the two regimes can be driven by the energy $E=eV_{p}$ of the applied charge gate $V_{p}$. Operatively,   the transconductance ${\rm G}_{12}$ drops from values  exponentially close to $e^2/h$, for $|eV_{p}| <|\Gamma_p|$, to algebraically small values, for $|eV_{p}| \gtrsim |\Gamma_p|$, as shown by the solid curves in Fig.\ref{Fig-ETJ-setup}(b). Notice that the energy window, over which the transconductance  drop occurs, decreases with increasing the length~$L$ of the junction. Also, the drop exhibits an oscillatory behavior  around the energy $|\Gamma_p|$, as a result of electron interference from `back-scattering' at the two boundaries of the tunnel junction, and also depends on the smoothing length $\lambda$ of the tunneling amplitude, here chosen to be $\lambda=0.25 L$. Details about the numerical calculations can be found in the Appendix.

\begin{figure} 
\centering
\includegraphics[width=\columnwidth,clip]{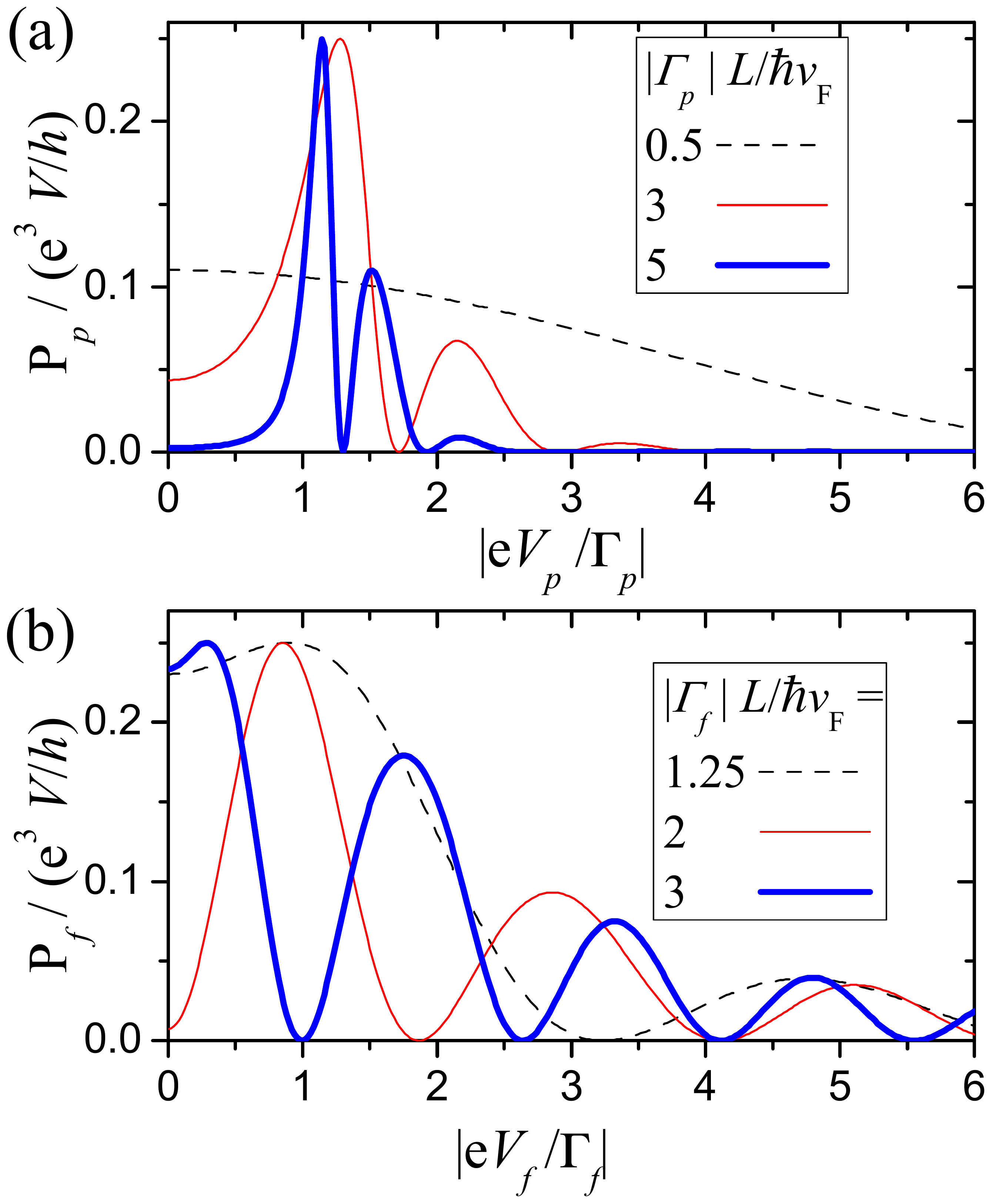}
\caption{(Color online) The two types of partition noise in the setup of Fig.\ref{Fig-ETJ-setup}. (a) The spin-preserving partition noise, detected from local noise measurement when the four terminals are set in the charge-bias configuration [see Eq.(\ref{P-p-def})] can be tuned with the charge gate voltage $V_{p}=(V_{T}+V_{B})/2$. For an ETJ ($|\Gamma_p| L/\hbar v_F>1$), it exhibits peaks  in the energy window of the crossover of the transconductance shown in Fig.\ref{Fig-ETJ-setup}(a).  (b) The spin-flipping partition noise Eq.(\ref{P-p-def}), detectable as local noise and as cross correlations when the four terminals are set in the spin-bias configuration and when the condition $R_{p}=0$ is fulfilled [vertical arrows in Fig.\ref{Fig-ETJ-setup}(b)], is plotted as a function of the spin gate voltage $V_{f}=(V_{T}-V_{B})/2$. The three curves refer to different values of the spin-flipping tunneling amplitude.}
\label{Fig-ETJ-noise}
\end{figure}

The existence of the transconductance  drop strongly determines the observability of the two types of partition noise. In the first instance, it is within the energy window of the drop that the spin-preserving partitioning $R_{p} T_{p}$ is non-vanishing, and can be detected from the local noise in the charge-bias configuration [see Eqs.(\ref{P-p-def})-(\ref{P-p-def-2})]. This is shown in Fig.\ref{Fig-ETJ-noise}(a), where the spin-preserving noise ${\rm P}_p$ is plotted as a function of the charge gate voltage~$V_{p}$.
Secondly, the transconductance drop  hallmarks the access to the regime where the condition $R_{p}\simeq 0$ is fulfilled. 
With maintaining $V_{p}$ within this range, highlighted by arrows in Fig.\ref{Fig-ETJ-setup}(b), the noise matrix in the spin-bias configuration acquires the form~(\ref{shot-SB-bis}), and the spin-flipping partition noise $R_{f} T_{f}$ can be directly measured both as local noise or cross correlation  between terminals (1,2) and (3,4)  [see Eq.(\ref{P-f-def})].
Notably, such noise can be controlled by sweeping the spin gate $V_{f}$, and exhibits oscillations that are related to the length $L$ of the junction, as shown in Fig.\ref{Fig-ETJ-noise}(b) for different values of the parameter $L |\Gamma_f|/\hbar v_F$.  \\

\subsection{Double Quantum Point Contact (DQPC)}
\begin{figure} 
\centering
\includegraphics[width=\columnwidth,clip]{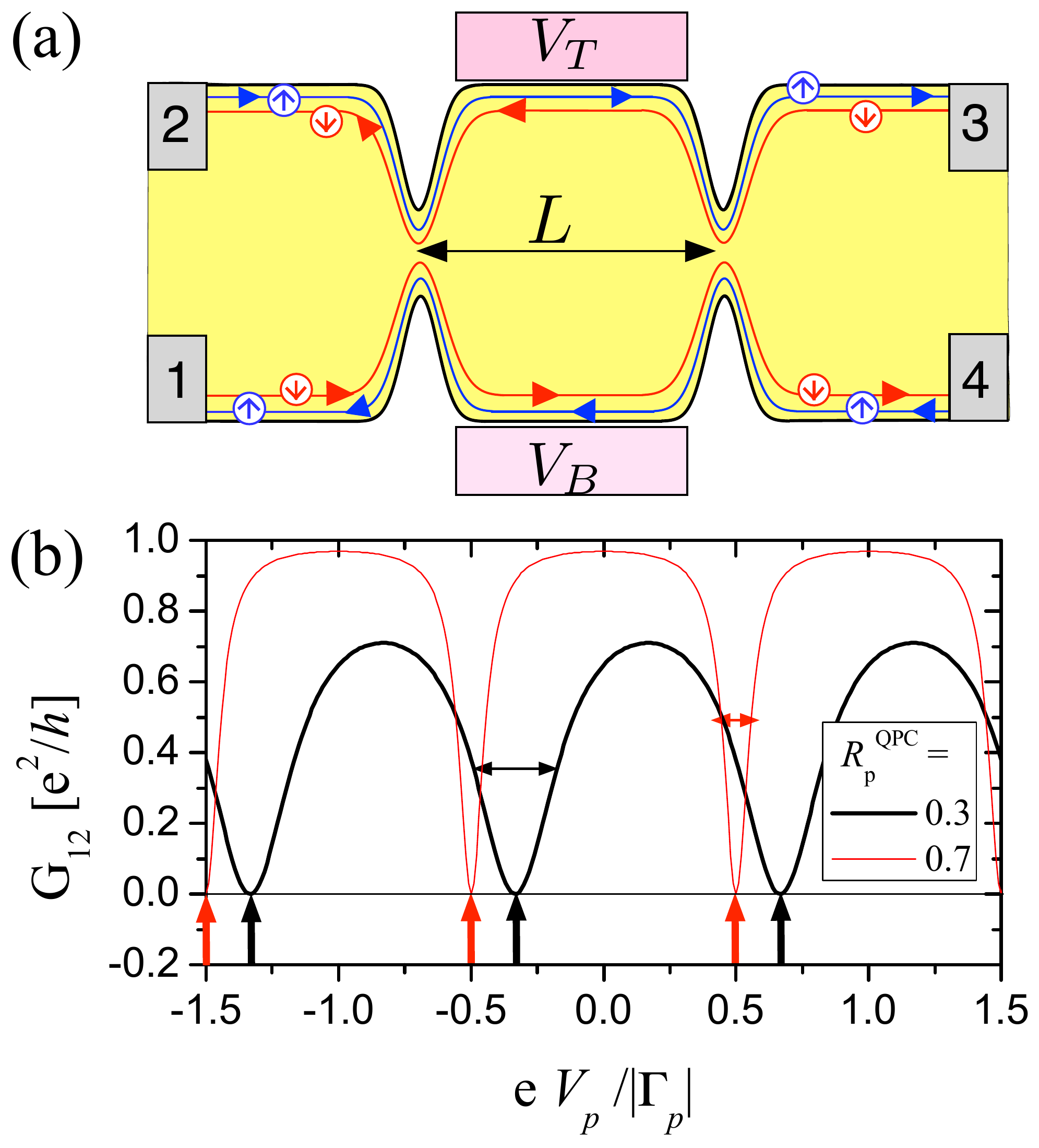}
\caption{(Color online) (a) Sketch of a double quantum point contact setup, where each of the two (equal)  QPC  is characterized by a spin-preserving and spin-flipping reflection coefficient $R_{p}^{\rm QPC}$ and $R_{f}^{\rm QPC}$, respectively. A phase difference $\Delta \phi_{p/f}$ is allowed to occur between the complex tunneling amplitudes $\Gamma_{p/f}$ of the two QPCs. 
(b) The transconductance $|{\rm G}_{12}|=(e^2/h) R_p$,  as a function of the charge gate voltage $V_{p}=(V_{T}+V_{B})/2$, for $R_{p}^{\rm QPC}=0.3$ and $\Delta \phi_{p}=\pi/3$ (thick black curve) and $R_{p}^{\rm QPC}=0.7$ and $\Delta \phi_{p}=0$ (thin red curve). The pattern  described by Eq.(\ref{RpDQPC}), exhibits a period $E_L= \pi \hbar v_F/L$ related to the distance $L$ between the QPCs, and  vanishing minima at values (\ref{Vpstar}), highlighted by vertical arrows. By setting $V_{p}$ to these values, the condition $R_{p}=0$ for the observation of the spin-flipping partitioning is fulfilled, see Eq.(\ref{P-f-def}).}
\label{Fig-DQPC-setup}
\end{figure}

The second proposed setup is depicted in Fig.\ref{Fig-DQPC-setup}(a), and consists of two short QPCs, separated by a distance~$L$. Although each QPC alone would yield a roughly  energy independent reflection $R_{p}^{\rm QPC}$, the interference between  tunneling events occurring at the two QPCs leads to an overall energy dependent reflection coefficient $R_{p}$ for the whole DQPC system. In this case,  analytic expressions can be  straightforwardly obtained by combining the transfer matrices of each QPC and the free propagation between them, as shown in Appendix. In particular, we shall consider the case where the two  QPCs have the same geometrical shape and are thus characterised by the same reflection coefficient $R_{p}^{\rm QPC}$. For the whole DQPC system one obtains (see Appendix for details)
\begin{equation}
R_{p}=1-\left[ 1+ \frac{4 R_{p}^{\rm  QPC}}{(1-R_{p}^{\rm QPC})^2} \cos^2(\frac{eV_{p} L}{\hbar v_F} -\frac{\Delta\phi_p}{2}) \right]^{-1} \label{RpDQPC}
\end{equation} 
where  $\Delta \phi_p$ accounts for a possible difference between the phases of the complex tunnelling amplitudes $\Gamma_p^{(1)}$ and $\Gamma_p^{(2)}$ at the two QPCs. Equation (\ref{RpDQPC}) describes a series of perfect resonances ($R_{p}=0$), which results into a periodic sequence of vanishing minima in the transconductance $|{\rm G}_{12}|=(e^2/h) R_p$ as a function of $V_{p}$, occurring at charge voltage values
\begin{equation}
e V_{p}^{*}=E_L \left(   n  +\frac{1}{2} +\frac{\Delta\phi_p}{2\pi} \right) \hspace{0.5cm} n=0,1,\ldots \quad, \label{Vpstar}
\end{equation}
shown  by the vertical arrows in Fig.\ref{Fig-DQPC-setup}(b). The energy scale $E_L= \pi \hbar v_F/L$ related to the distance 
$L$ between the two QPCs determines the pattern period. For each minimum, the full-width at half-minimum  depends on the value of the spin-preserving reflection $R_{p}^{\rm QPC}$ of each individual QPC  
\begin{equation}
\Delta(e V_{p})=\frac{2\hbar v_F}{L} \arcsin\left[\left( 1+\left(\frac{1+R_{p}^{\rm QPC}}{1-R_{p}^{\rm QPC}}\right)^2\right)^{\!-\frac{1}{2}} \right] 
\end{equation}
and is   highlighted by the horizontal arrows in Fig.\ref{Fig-DQPC-setup}(b).\\

\begin{figure} 
\centering
\includegraphics[width=\columnwidth,clip]{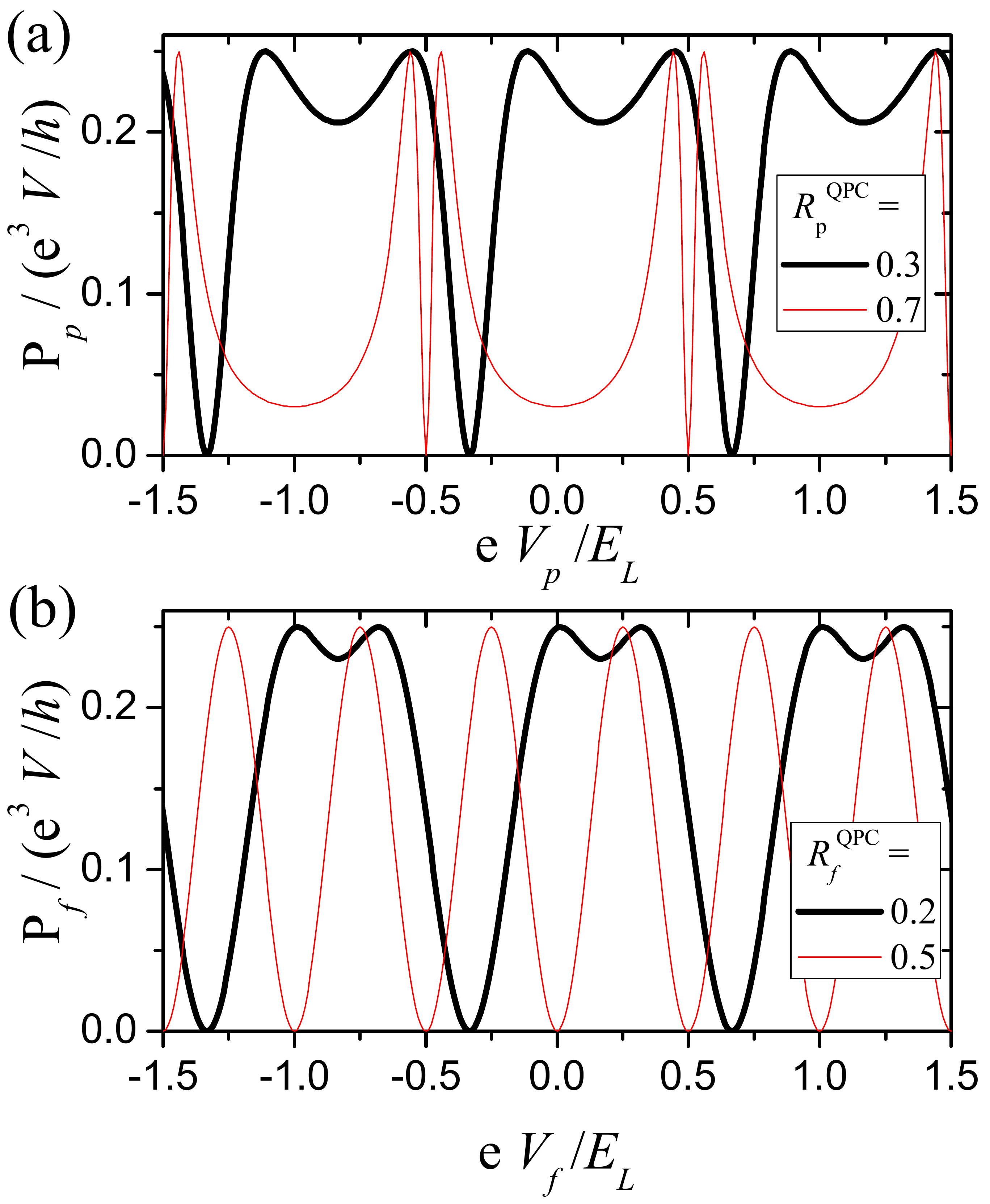}
\caption{(Color online) The two types of partition noise in the DQPC setup of Fig.\ref{Fig-DQPC-setup}. (a) The spin-preserving partition noise, detected from local noise measurement when the four terminals are set in the charge-bias configuration [see Eq.(\ref{P-p-def})] as a function of the charge gate voltage $V_{p}=(V_{T}+V_{B})/2$. The thick black and the thin red curves refer to parameters $R_{p}^{\rm QPC}=0.3$, $\Delta \phi_{p}=\pi/3$ and $R_{p}^{\rm QPC}=0.7$, $\Delta \phi_{p}=0$, respectively. (b) The spin-flipping partition noise Eq.(\ref{P-p-def}), detectable as local noise and as cross correlations when the four terminals are set in the spin-bias configuration and when the condition $R_{p}=0$ is fulfilled (vertical arrows in Fig.\ref{Fig-DQPC-setup}(b)), is plotted as a function of the spin gate voltage $V_{f}=(V_{T}-V_{B})/2$. The thick black and the thin red curves refer to the parameters $R_{f}^{\rm QPC}=0.2$, $\Delta \phi_{f}=\pi/3$ and $R_{f}^{\rm QPC}=0.5$, $\Delta \phi_{f}=0$, respectively.}
\label{Fig-DQPC-noise}
\end{figure}

The corresponding spin-preserving partition noise ${\rm P}_p$ also exhibits a periodic pattern as a function of $V_{p}$, with nodes at the values (\ref{Vpstar}), as shown in Fig.\ref{Fig-DQPC-noise}(a) for two different values of the single QPC reflection $R_{p}^{\rm QPC}$.
By adjusting the charge bias at the  values (\ref{Vpstar}), the spin-flipping partitioning $R_{f} T_{f}$
can be directly measured both as local noise and as cross correlations between terminals (1,2) and (3,4), as described by Eq.(\ref{shot-SB-bis}). In this case the DQPC spin-flipping reflection is given by
\begin{equation}
R_{f}=1-4 R_{f}^{\rm QPC}(1-R_{f}^{\rm QPC}) \cos^2 \left( \frac{eV_{f}}{\hbar v_F}-\frac{\Delta \phi_f}{2} \right)
\end{equation}
with $R_{f}^{\rm QPC}$ denoting the spin-flipping reflection of each individual QPC.
Similarly to the case of the ETJ, such noise can be controlled by sweeping the spin gate $V_{f}$, as shown in Fig.\ref{Fig-DQPC-noise}(b) for different values of the single-QPC spin-flipping transmission. In the DQPC setup, however, the noise oscillations   are related to the distance~$L$ between the two QPCs.  

\section{Discussion and Conclusions}
\label{sec-V}
We have investigated the current-current correlations~${\rm P}_{ij}$ in a four-terminal QSH setup, where the tunnel  coupling between helical edge states involves both spin-preserving and spin-flipping  processes, as a consequence of the spin-orbit interaction characterizing these systems. 

In Sec.~\ref{sec-III} general results have been presented:
First, focussing on the Johnson-Nyquist correlations at thermal equilibrium, Eq.(\ref{P-JN}), we have shown that the local noise is independent of the features of the tunnel region, and is given by the universal expression ${\rm P}_{ii}=2 k_B T e^2/h$, where~$T$ is the temperature. This behavior is a hallmark of the topological protection from backscattering in helical edge states. In contrast, the off-diagonal entries ${\rm P}_{i \neq j}$ enable one to operatively determine the reflection coefficients $R_{p}$ and $R_{f}$ of the two tunneling processes, via the cross-correlations between different pairs of terminals. This detection scheme is equivalent and alternative to transconductance measurements.

Secondly, we have analyzed the richer scenario of out-of-equilibrium noise $eV \gg k_B T$, where various voltage bias configurations for the four terminals are possible. When tunneling is absent, the current-current correlations vanish, showing that decoupled helical edge states are noiseless at low temperatures. However, in the presence of tunneling, current correlations do appear, and each entry ${\rm P}_{ij}$ consists of mixed products of transmission and reflection involving both spin-preserving and spin-flipping processes [see Eqs.(\ref{shot-CB})-(\ref{shot-SB})]. Nevertheless, we have identified the conditions for a direct detection of the two types of partitioning related to these  processes.  We have shown that, while the spin-preserving partitioning $R_{p} T_{p}$ can be directly probed for any tunnel junction via a measurement of the local noise  ${\rm P}_{ii}$ by adopting a charge-bias configuration for the terminals  [see Eqs.(\ref{P-p-def}) and (\ref{P-p-def-2})], the spin-flipping partitioning $R_{f} T_{f}$ can be detected only under the condition $R_{p}=0$ and $R_{f} \neq 0$, in the spin-bias configuration [see Eq.(\ref{P-f-def})]. 

In Sec.~\ref{sec-IV}, we have shown that such condition can be fulfilled in suitably designed setups, by exploiting the energy dependence of reflection coefficient related to the $p$-process. Two proposals of such  setups have been discussed, namely, the ETJ and the DQPC, which act as a high-pass energy filter and an electron interferometer, respectively, as can be seen from the behavior of their  ${\rm G}_{12}$ transconductance  (Figs.\ref{Fig-ETJ-setup}  and \ref{Fig-DQPC-setup}). In these cases both partition noises ${\rm P}_p$ and ${\rm P}_f$ can be detected and controlled independently by   combinations $V_{p/f}=(V_{T}\pm V_{B})/2$ of suitable gate voltages.   For the ETJ setup, ${\rm P}_p$ is localized over the energy window related to the length of the junction [see Fig.\ref{Fig-ETJ-noise}(a)] and, by setting~$V_{p}$ beyond such energy window, the spin-flipping partitioning ${\rm P}_f$ can be detected both as local noise in any terminal and as cross-correlations between terminal pairs  (1,2)  and  (3,4), in the spin-bias configuration [see Fig.\ref{Fig-ETJ-noise}(b)]. 
For the DQPC setup, ${\rm P}_p$ exhibits a  pattern characterized by a period $E_L$ related to the distance between the two point-contacts, and by a width of the minima related to the reflection of each individual QPC [see Fig.\ref{Fig-DQPC-noise}(a)], in agreement with the results of Ref.[\onlinecite{moskalets_2013}]. By setting~$V_{p}$ to the minima values (\ref{Vpstar}), the spin-flipping partition noise can be straightforwardly detected and tuned with the spin gate~$V_{f}$,  within the spin-bias configuration of the terminals [see Fig.\ref{Fig-DQPC-noise}(b)].\\

Finally, we observe that tunnel junctions in QSH bars can be realized by lateral etching of the quantum well, and the above setups can be tailored with standard lithographic techniques. Lateral gates $V_{T}$ and $V_{B}$ can be also  implemented. 
In HgTe/CdTe quantum wells, a $100-200 \, {\rm nm}$ wide constriction   leads to a tunneling amplitude magnitude $|\Gamma_p| \sim 0.25 - 2.5 \, {\rm meV}$,\cite{zhou,richter}  well below the bulk gap. Using the value of the Fermi velocity $v_F \simeq 0.5 \times   10^6 \, {\rm m/s}$,~\cite{molenkamp-zhang_jpsj}  a  lengthscale  $\xi_0=\hbar v_F/|\Gamma_p| \sim 0.1 - 1 \,\mu{\rm m}$ is obtained. The ETJ regime can thus be realized with   a $L \sim 1 \mu {\rm m}$ long constriction. For InAs/GaSb quantum wells, the conditions are even more favorable, since the smaller Fermi velocity $v_F \simeq 2 \times   10^4 \, {\rm m/s}$ leads to shorter values for $\xi_0$.~\cite{liu-zhang_2008,knez_2014} The value of the spin-flipping magnitude  $|\Gamma_f|$ is typically smaller than $|\Gamma_p|$ by a factor 3 to 4 only.~\cite{zhou,richter,sternativo1,citro-sassetti,moskalets_2013} 
Then, within the suitable voltage bias configurations discussed above, the two partition noises ${\rm P}_p$ and $P_{f}$ can be tuned with varying the charge and spin gate voltages $V_{p}$ and $V_{f}$ in the ${\rm meV}$ regime.   Similarly, in the DQPC setup the individual PQC reflection is controlled by the constrictions, and a distance of about $1 \mu {\rm m}$ between the two QPCs  leads to a pattern for $P_{p}$ and $P_{f}$ that has a typical period of a few ${\rm meV}$. The possibility to detect the two partition noises related to spin-preserving and spin-flipping processes thus seems to be at experimental reach.

\acknowledgments
The author  greatly acknowledges   support from Italian FIRB 2012 project HybridNanoDev (Grant No.RBFR1236VV).

\appendix*
\section{Technical aspects about the ETJ and DQPC setups}
In Sec.~\ref{sec-III} we have shown that each current-current correlation entry ${\rm P}_{ij}$ consists of a product of various combinations of the reflection and transmission coefficients related to $p$- and $f$-processes [see Eqs.(\ref{shot-CB}) and (\ref{shot-SB})]. In this Appendix we provide  technical details about the computation of the transmission coefficients $T_{p}$ and $T_{f}$ for the ETJ and DQPC setups proposed in Sec.~\ref{sec-IV}. The reflection coefficients are then straightforwardly obtained as $R_{p/f}=1-T_{p/f}$. 

We start by describing the general computation scheme.
As shown in Refs.[\onlinecite{sternativo1},\onlinecite{sternativo2}], for a tunnel region characterized by   arbitrary profiles of tunneling amplitudes $\Gamma_{p}(x), \, \Gamma_f(x)$ and charge and spin gate voltages $V_{p}(x), \, V_f(x)$,
the transfer matrix ${\mathsf{M}}$ of the four-terminal setup turns out to be a direct product of spin-preserving and spin-flipping components, $
{\mathsf{M}}=  \mathbf{m}_f  \,   \otimes \, \mathbf{m}_p   
$, which determine the transmission coefficients  related to $p$- and $f$- tunneling processes  through the relations~\cite{sternativo1}
\begin{eqnarray}
T_{p}&=&  |(\mathbf{m}_p)_{22}|^{-2}  \label{Tp-gen}\\
T_{f}&=&   |(\mathbf{m}_f)_{22}|^2 \label{Tf-gen}  \quad,
\end{eqnarray} 
where
\begin{eqnarray}
\mathbf{m}_p & {\,=\,} &  e^{-i \tau_z k_E x_f}  \mathbf{U}_{p}(x_f;x_0)  e^{+i \tau_z k_E x_0} \, \, \, \label{mp-def}\\
\mathbf{m}_f & {\,=\,} & \mathbf{U}_f(x_f;x_0) \quad. \label{mf-def} 
\end{eqnarray}
In Eqs.(\ref{Tp-gen}) and (\ref{Tf-gen}), $x_0$ and $x_f$ are the extremal longitudinal coordinates of the tunnel region, on the left and on the right, respectively. Furthermore
\begin{eqnarray}
\mathbf{U}_{p}(x;0)&= &  {\rm T} \,\exp\left[{\displaystyle -i \int_{0}^x \, dx^\prime  \boldsymbol\tau  \cdot \mathbf{b}_{p,E}(x^\prime) \,} \right]\label{Up-expr} \\
\mathbf{U}_f(x;0)&= &   {\rm T} \,\exp\left[{\displaystyle -i \int_{0}^x \, dx^\prime  \boldsymbol\sigma   \cdot \mathbf{b}_f(x^\prime)   }\right] \label{Uf-expr} \end{eqnarray}
are ``evolution'' operators (in space), with $\boldsymbol\sigma=(\sigma_x,\sigma_y,\sigma_z)$ and $\boldsymbol\tau=(\tau_x,\tau_y,\tau_z)$ denoting Pauli matrices in spin ($\sigma=\uparrow, \downarrow$) and chirality ($\alpha=R,L$) space, respectively, and
\begin{equation}
 \label{bE-def}
\mathbf{b}_{p,E}(x) =   (-i\, {\rm Im}  \Gamma_{p}(x)   \, , \,  i\, {\rm Re}  \Gamma_{p}(x)     , \,  eV_{p}(x) -E) / \hbar v_F  
\end{equation}
\begin{eqnarray}
\mathbf{b}_f(x)= \left({\rm Re}  \Gamma_{f}(x)  ,  {\rm Im} \Gamma_{f}(x) ,  eV_{f}(x) \right) / \hbar v_F \hspace{1cm} \label{bf-def}
\end{eqnarray}
involving spin-preserving and spin-flipping tunneling amplitudes, and charge and spin gate voltages, respectively.

Any generic profile $\Gamma_{\nu}(x),V_{\nu}(x)$ ($\nu=p,f$) can be treated with arbitrarily high precision by dividing the tunnel region into a sufficiently high number $N$ of equal intervals $[x_{j-1};x_{j}]$ ($j=1,\ldots N$), where the profile is approximated by a locally constant value  
\begin{eqnarray}
\left\{ \begin{array}{lcl}
 \Gamma_\nu(x) & \equiv  &\Gamma_\nu^{(j)}  \\
 & &   \\
V_\nu(x) & \equiv  &V_\nu^{(j)}  
\end{array} \right. \hspace{1cm} x_{j-1} \le  x < x_j
 \quad. \label{loc-const}
\end{eqnarray} 
The total evolution operator across the junction is then given by the product
\begin{eqnarray}
\mathbf{U}_{\nu}(x_f;x_0)  = \prod_{j=N}^1 \mathbf{U}_{\nu}(x_j;x_{j-1}) \hspace{0.5cm}  \label{U-c-uG_1}  \nu=p,f\quad,  
\end{eqnarray}
 of short evolutions over the locally constant profiles, which can be straightforwardly computed.~\cite{sternativo1} One obtains for the $p$-sector
\begin{eqnarray}
\lefteqn{\mathbf{U}_{p}(x_{j};x_{j-1}) =} \hspace{1cm} & &  \nonumber \\ 
 & &   \,    \nonumber    \\
& & =  \left\{ \begin{array}{l}
 \tau_0 \cos(\tilde{k}_E^{(j)} {l})\,  -i  \boldsymbol\tau \cdot \mathbf{b}_{p,E}^{(j)}   \frac{\sin(\tilde{k}_E^{(j)} {l}) }{\tilde{k}_E^{(j)}}      \,  \\ \\  \hspace{3.5cm} \mbox{for} \, \,  |E|>|\Gamma_{p}^{(j)}| \\ \\ \\
 \tau_0\cosh(\tilde{q}_E^{(j)}  {l}) \,  - i \boldsymbol\tau \cdot \mathbf{b}_{p,E}^{(j)}  \frac{\sinh(\tilde{q}^{(j)}_E{l}) }{ \tilde{q}_E^{(j)} }   \, \\ \\
   \hspace{3.5 cm} \mbox{for} \, \, |E|<|\Gamma_{p}^{(j)}| \end{array}\right.
 \label{Up-const}
\end{eqnarray}
with
$\tilde{k}^{(j)}_E =  \sqrt{ E^2-|\Gamma_{p}^{(j)}|^2}/\hbar v_F$  for $|E|>|\Gamma_{p}^{(j)}$ and $\tilde{q}_E^{(j)} = \sqrt{|\Gamma_{p}^{(j)}|^2-E^2}/\hbar v_F$  for $|E|<|\Gamma_{p}^{(j)}|$,  
$\mathbf{b}^{(j)}_{p,E}   =   (-i\,  {\rm Im} \Gamma_{p}^{(j)}    , \,  i\, {\rm Re} \Gamma_{p}^{(j)}    \, , eV_{p}^{(j)}\,   -E)/\hbar v_F$, 
and for the $f$-sector  
\begin{eqnarray}
\mathbf{U}_{f}(x_{j};x_{j-1}) =  \sigma_0 \cos(\tilde{k}_{f}^{(j)} l) -i  \boldsymbol\sigma \cdot  \mathbf{b}_f \frac{\sin(\tilde{k}_{f}^{(j)} l)}{\tilde{k}_{f}^{(j)}} \hspace{1cm} 
\label{Uf-const}
\end{eqnarray}
with
$
\tilde{k}^{(j)}_{f} = \sqrt{|\Gamma_{f}^{(j)}|^2+(e V_{f}^{(j)})^2} \, /\hbar v_F 
$. In Eqs.(\ref{Up-const})-(\ref{Uf-const}) $l=L/N$ denotes the length of each interval.\\

The above procedure can be applied to describe any tunnel junction setup. Let us now focus on the specific setups presented in Sec.~\ref{sec-IV}.

In particular, for the ETJ setup, the  profile assumed for the magnitude of the tunneling amplitude, depicted in Fig.\ref{Fig-ETJ-setup}(a), is
\begin{equation}
|\Gamma_\nu(x)| \doteq |\Gamma_\nu| f( x +\frac{L}{2}) \, f(-x+\frac{L}{2}) \hspace{0.5cm} \nu=p,f \label{Gamma-ETJ}
\end{equation}
where $|\Gamma_\nu|$ is the bulk value, and
\begin{equation}
f(x) \doteq  \frac{1}{2} \left(1+\tanh\left(\frac{5}{2\lambda}(x+\frac{\lambda}{2})\right) \right)\quad.
\end{equation}
is a smoothening function that interpolates between 0 and 1 over a distance $\lambda$.  By dividing the tunneling region into $N \sim 10^2$ intervals, and by approximating (\ref{Gamma-ETJ}) with piecewise constant values (\ref{loc-const}), the operators (\ref{U-c-uG_1}) are built, and   the transmission coefficients (\ref{Tp-gen}) and (\ref{Tf-gen}) are obtained from Eqs.(\ref{mp-def}) and (\ref{mf-def}), respectively.

For the DQPC setup, the total evolution operators (\ref{Up-expr})-(\ref{Uf-expr}) can be constructed by combining the evolution across the two QPCs with the free evolution  between them. Each QPC is a short junction where $l_{QPC} \rightarrow 0$, $|\Gamma_\nu|\rightarrow \infty$, with the parameter  $a_\nu=|\Gamma_\nu|l_{QPC}/\hbar v_F$ kept   constant. Then Eqs.(\ref{Up-const}) and (\ref{Uf-const}) reduce to 
\begin{equation}
\mathbf{U}_{p}^{\rm QPC}= \left(\begin{array}{ccc} \cosh{a_p} & -i \sinh{a_p} \, e^{-i\phi_p} \\ & \\i  e^{i\phi_p} \sinh{a_p} \, & \cosh{a_p} 
\end{array} \right) 
\end{equation}
and
\begin{equation}
\mathbf{U}_{f}^{\rm QPC}= \left(\begin{array}{ccc} \cos{a_f} & -i  e^{-i\phi_f}\sin{a_f} \, \\ & \\-i  e^{i\phi_f} \sin{a_f} \, & \cos{a_f} 
\end{array} \right) \, \, ,
\end{equation}
where $\phi_\nu$ ($\nu=p,f$) is the phase of the complex tunneling amplitude $\Gamma_\nu=|\Gamma_\nu| e^{i \phi_\nu}$.
Then, the total DQPC evolution operators are simply built as 
\begin{equation}
\mathbf{U}_p(x_f;x_0)=\mathbf{U}_{p}^{\rm QPC_2} \, e^{i (E-eV_p) L \tau_z/\hbar v_F} \, \mathbf{U}_{p}^{\rm QPC_1} \label{Up-DQPC}
\end{equation}
and
\begin{equation}
\mathbf{U}_f(x_f;x_0)=\mathbf{U}_{f}^{\rm QPC_2} \, e^{-i  eV_f  L \sigma_z/\hbar v_F} \, \mathbf{U}_{f}^{\rm QPC_1} \quad. \label{Uf-DQPC}
\end{equation}
Inserting Eqs.(\ref{Up-DQPC}) and (\ref{Uf-DQPC}) into Eqs.(\ref{mp-def}) and (\ref{mf-def}), respectively, the transmission coefficients (\ref{Tp-gen}) and (\ref{Tf-gen}) are then obtained.

\end{document}